\documentclass{article}

\usepackage{arxiv}

\usepackage[utf8]{inputenc} 
\usepackage[nottoc]{tocbibind}
\usepackage[T1]{fontenc}    
\usepackage{hyperref}       
\usepackage{url}            
\usepackage{booktabs}       
\usepackage{amsfonts}       
\usepackage{nicefrac}       
\usepackage{microtype}      
\usepackage{cite}
\usepackage[table]{xcolor}
\usepackage{amsmath,amssymb,amsfonts}
\usepackage{algorithmic}
\usepackage{graphicx}
\usepackage{textcomp}
\usepackage{xcolor}
\usepackage{hyperref}
\usepackage{nicefrac}
\usepackage{paralist}
\usepackage{algorithmic}
\usepackage{graphicx}
\usepackage{textcomp}
\usepackage[table]{xcolor}
\usepackage{balance}
\usepackage{enumitem}
\usepackage{booktabs}
\usepackage{multirow}
\usepackage{hhline}
\usepackage{url}
\usepackage{tcolorbox}
\usepackage{xspace}
\usepackage{stfloats}
\def\BibTeX{{\rm B\kern-.05em{\sc i\kern-.025em b}\kern-.08em
    T\kern-.1667em\lower.7ex\hbox{E}\kern-.125emX}}
\usepackage{paralist}

\title{Towards Continuous Systematic Literature Review in Software Engineering}

\author{
Bianca Minetto Napole\~ao\\
Université du Québec à Chicoutimi \\
Chicoutimi, QC, Canada \\
\texttt{bianca-minetto.napoleao1@uqca.ca} 
\And
Fabio Petrillo\\
École de Technologie Supérieure  \\
Montreal, QC, Canada\\
\texttt{fabio@petrillo.com} 
\And
Sylvain Hall\'e\\
Université du Québec à Chicoutimi \\
Chicoutimi, QC, Canada \\
\texttt{shalle@acm.org} 
\And
Marcos Kalinowski\\
Pontifical Catholic University of Rio de Janeiro\\
Rio de Janeiro, RJ, Brazil \\
\texttt{kalinowski@inf.puc-rio.br} 
}

\date{}
\begin{document}
\maketitle

\begin{abstract}
New scientific evidence continuously arises with advances in Software Engineering (SE) research. Conventionally, Systematic Literature Reviews (SLRs) are not updated or updated intermittently, leaving gaps between updates, during which time the SLR may be missing crucial new evidence. 
\textbf{Goal:} We propose and evaluate a concept and process called Continuous Systematic Literature Review (CSLR) in SE.
\textbf{Method:} To elaborate on the CSLR concept and process, we performed a synthesis of evidence by conducting a meta-ethnography, addressing knowledge from varied research areas. Furthermore, we conducted a case study 
to evaluate the CSLR process.
\textbf{Results:} We describe the resulting CSLR process in BPMN format. The case study results provide indications on the importance and feasibility of applying CSLR in practice to continuously update SLR evidence in SE.
\textbf{Conclusion:} The CSLR concept and process provide a feasible and systematic way to continuously incorporate new evidence into SLRs, supporting trustworthy and up-to-date evidence for SLRs in SE.
\end{abstract}


\section{Introduction}
\label{sec:introduction}

Over the last ten years, Systematic Literature Reviews (SLRs) updates have been conducted in the Evidence-Based Software Engineering (EBSE) area.
According to Mendes \textit{\textit{et al.} } \cite{Mendes2020}, an SLR update is a more recent (updated) version of an SLR that includes new evidence (primary studies). They may also include new methods such as new quality criteria to evaluate evidence, different search strategies to detect new evidence and repeat or remake (using a more recent synthesis method, for example) the analysis of the original SLR.  

One known challenge in evidence-based disciplines is to keep SLRs updated. As stated in Cochrane's handbook \cite{Cochraine2019a}, an SLR that is not maintained may become out-of-date or misleading. Furthermore, with the advance of the computer science field and the growth of research publications, new evidence continuously arises. This fact impacts directly on the purpose of keeping SLRs up to date. An SLR that is out-of-date could lead researchers to obsolete conclusions or decisions about a research topic \cite{Watanabe20}. 

In the medicine field, SLR update has a consolidated process \cite{MedMoher2008}. Despite the effort of the Software Engineering (SE) community to keep SLRs updated, a recent study \cite{Mendes2020} showed that by 2019 only 20 SLRs in SE were updated since the start of SLR publication in 2004. In February 2021, the SE area had more than 1000 secondary studies (SLRs and Systematic Mappings (SMs)) published in several SE venues \cite{Napoleao2021S}. 

Creating and maintaining up-to-date SLRs demands a significant effort for reasons such as the rapid increase in the amount of evidence  and the limitation of available databases \cite{Zhang18}. 
Furthermore, the lack of detailed protocol documentation and data availability 
makes the SLR update process even more difficult since most of the tacit knowledge from the SLR conduction is lost \cite{Felizardo20}.

Conventionally, SE SLRs are not updated or updated intermittently \cite{Wohlin2020}. Periodically updating leaves gaps between updates, during which time the SLR may be missing crucial new research, placing it at risk of being inaccurate and wasting the potential contribution of new research to evidence synthesis and decision-making. Research efforts are needed to remedy these knowledge gaps in updating SLRs and ascertain the potential benefits of continuously assessing new evidence and evaluating the need of updating SLRs.

Inspired by the Living Systematic Review (LSR) \cite{Elliott17} idea from 
medicine and considering the DevOps concept \cite{Bass2015} from software development, especially the Continuous Integration and Delivery (CI/CD) processes and practices as well as OS practices, in this study, we propose the concept and process of Continuous Systematic Literature Review (CSLR) in SE aiming to support keeping SLRs constantly updated. Therefore, we compared, analyzed, interpreted and translated findings of studies through the conduction of a meta-ethnography \cite{Noblit/1988}. We also conducted a case study to evaluate the feasibility of applying the proposed CSLR concept and process. 

The contributions of this study are: (i) the proposition of the CSLR in SE concept; (ii) the CSLR process definition from the meta-ethnography conduction; and (iii) a case study to evaluate the feasibility of the CSLR process. Our main findings indicate that the CSLR concept and process provide an innovative and systematic way that can be applied to help maintaining SLRs, supporting continuously considering trustworthy and up-to-date evidence for SLRs in SE.

The remainder of this paper is organized as follows. Section \ref{sec:relatedwork} presents the related work. Section \ref{sec:towardscslr} describes the study design and the meta-ethnography conduction. 
Section \ref{sec:validation} presents the case study. 
Section \ref{sec:discussion} discusses our findings and highlights potential implications. The threats to validity are described in Section \ref{sec:threats}. Section \ref{sec:conclusion} concludes our study.

\section{Related Work} 
\label{sec:relatedwork}

In SE, there are initiatives on SLR updates such as on formalizing an SLR update process \cite{Dieste08a}, searching for new/updated evidence \cite{felizardo16, Wohlin2020}, selecting updated evidence \cite{Watanabe20}, deciding on whether to update or not \cite{Mendes2020}, and experience reports \cite{Garces17, Felizardo20}. However, searching for new evidence, selecting evidence, deciding upon updating, and enacting the update process following lessons learned from experience reports are important pieces of a problem that has not yet been explicitly addressed as a whole: the problem of leaving gaps between the SLR publication and possible updates.


A recent study \cite{Santos2021} proposes a sustainability view for SLR and SLR updates. It states the need for SE community efforts to promote the aspects: (i) social -- researcher's communication and participation during the SLR conduction and update; (ii) economic -- resources reduction during the SLR conduction and update; and (iii) technical -- supporting tools and technologies to conduct and update SLR. Our study proposes and evaluates a process that considers these three aspects. 

\section {Towards CSLR}
\label{sec:towardscslr} 

In medicine, in order to mitigate the SLR updating issue, Elliot \textit{et al.}  \cite{Elliott17} introduced the concept of Living Systematic Review (LSR). An LSR is an SLR that is continually updated, incorporating relevant new evidence as it becomes available. 

In software engineering, DevOps concepts (mindset, practices, and tools) were introduced to promote several benefits, such as the faster release of features, improved monitoring systems in production, stimulating collaboration among team members, and others \cite{Bass2015}. In addition, the CI DevOps practice aims to build and integrate all working versions of the software code, keeping the software updated. It includes automation, for example, a building service, and a cultural mindset such as integrating changes constantly \cite{Humble2010}. Maintaining an SLR up to date requires protocol changes, searching for new evidence, and management strategies to support the demand for updating. Thus, concepts inspired in DevOps concepts could represent an alternative to support continuous changes in SLRs to keep them up to date and with reliable results.

Moreover, an important initiative in the SE research community is promoting Open Science (OS). It consists of making research artifacts available to the public addressing open access, open data, and open-source practices \cite{Mendez2020}. OS  practices can directly impact SLR conduction, not limited to the access and availability of primary evidence for the conduction of SLRs, but in their adoption during the conduction of SLRs that reflect on the reproducibility and maintainability of the SLRs. Thus, researchers should address OS  practices in any EBSE study.

To address the problem of leaving gaps between the SLR publication and possible updates, we introduce the CSLR concept. CSLR comprises a continuous and systematic surveillance and analysis of potential new relevant evidence for published SLRs, contributing to keeping SLRs up to date. We designed a CSLR process by applying meta-ethnography, considering elements of the traditional SE SLR process \cite{Kitchenham07}, concepts from studies addressing supporting activities involved in updating SLRs in SE \cite{Wohlin2020, Mendes2020, Felizardo14}, LSR from medicine \cite{Elliott17}, DevOps \cite{Humble2010, Bass2015} and OS practices in SE \cite{Mendez2020}.


\subsection{Study Design}

\begin{figure} [ht!]
     \centering
     \includegraphics [width=0.6\linewidth]{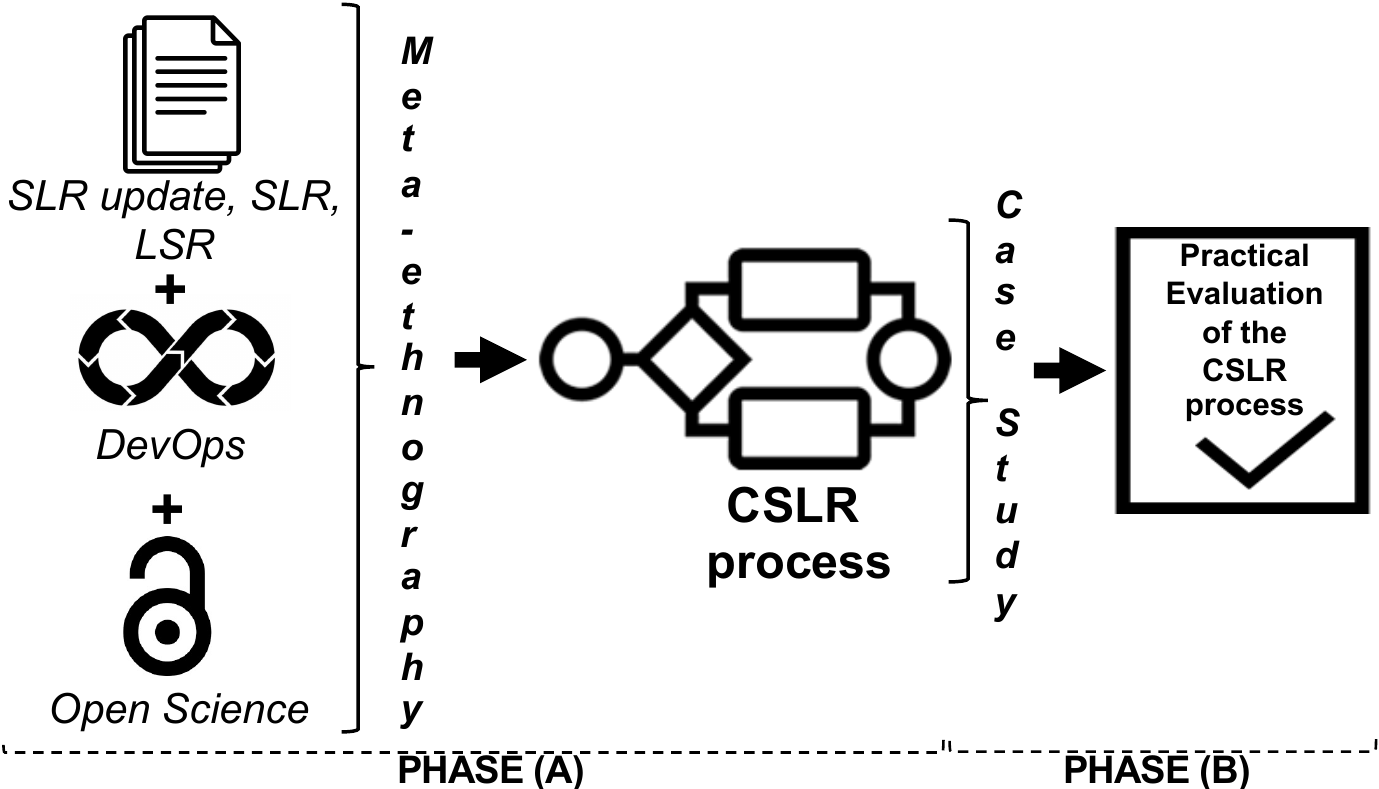}
     \caption{Study design summary}
     \label{fig:studydesignsummary}
 \end{figure}

In order to reach and validate our study goals, we divided it into two phases: (A) Creation of the CSLR process; and (B) Evaluation of the CSLR process. Figure \ref{fig:studydesignsummary} illustrates a summary of our study design. 

The research method adopted in phase (A) is meta-ethnography \cite{Noblit/1988} since it enables a systematic qualitative synthesis and detailed understanding of how topics and studies are related. According to Hannes \& Lockwood \cite{Hannes2012}, the meta-ethnography method provides a picture of the whole phenomenon under investigation from studies and its parts. In this sense, the resulting picture of our meta-ethnography is the CSLR process and its steps. The meta-ethnography method, in summary, involves researchers selecting, analyzing and interpreting qualitative studies through a process of translation, which provides an interpretation of the entire topic, in order to answer questions, gain new insights and/or build knowledge on a specific topic \cite{Noblit/1988}.

During phase (B) we performed a case study \cite{Runeson12} with a well-known SLR in SE to evaluate the CSLR process observing its contributions to mitigating the SLR intermittent update issues. The details about the conduction of each phase are described in Sections \ref{subsec:metaethnography} and \ref{sec:validation} respectively. 

\subsection{Application of the meta-ethnography method}
\label{subsec:metaethnography}
We present the seven stages of the meta-ethnography \cite{Noblit/1988} with a brief description and the application results from each of the stages. 

\subsubsection{Getting started}

During this first stage, the goal is to identify a topic of interest to be qualitatively explored and then define a Research Question (RQ) that represents the topic and guides the research \cite{Noblit/1988}. In this study, we aim to understand how concepts from the SLR process, SLR update, LSR, DevOps, and OS are related and how they can be integrated to mitigate intermittent SLR update issues in SE. 


{\textit{RQ: How are the SLRs process, SLR update, LSR, DevOps and OS concepts related and how can they be integrated to help mitigate intermittent SLR update issues in SE?}}


To answer our RQ, we synthesized intersections and relationships of these concepts to create the CSLR process.

\subsubsection{Deciding what studies are relevant to the topic of interest}

In the second stage, the goal is to find and select relevant studies to the topic of interest. It includes searching and selecting studies to be analyzed \cite{Noblit/1988}. Firstly, we performed a systematic search on Scopus using the terms: \textit{((``systematic literature review process''  OR  ``systematic review process''  OR  ``systematic literature review guideline''  OR  ``systematic review update''  OR  ``SLR Update'')  AND  ``software engineering'')} to detect studies that address the SLR process and SLR update strategies. Secondly, we searched for studies and guidelines that addresses LSR on Scopus using the the term: \textit{(``living systematic review'')} and also in two renown medicine databases:  PubMed and Cochrane.
We performed both searches in February 2022. Thirdly, since our goal is to synthesize information from DevOps practices, we opted to use as a base for our analysis two widely adopted books that describe the DevOps process \cite{Bass2015} and the CI/CD pipeline \cite{Humble2010}. Finally, for OS practices, we considered the book chapter of Mendez \textit{et al.} \cite{Mendez2020} since it presents a recent overview of OS practices in SE.

In order to select relevant studies on SLR, SLR update and LSR, we defined three Inclusion Criteria (IC) and three Exclusion Criteria (EC): (IC1) The study proposes or discusses a process or elements of a process on SLR Update in SE or LSR in medicine; \textit{OR} (IC2) The study is a guideline for the conduction of SLR, SLR Update in SE or LSR in medicine; \textit{OR} (IC3) The study is an experience report on SLR update in SE. (EC1) The study is an SLR, SLR update or LSR, but it does not discuss any step or aspect of the SLR, SLR update and LSR processes; \textit{OR} (EC2) The study is an experience report on SLR conduction; \textit{OR} (EC3) The study is an older version of another study already considered; \textit{OR} (EC3) The study is not written in English. 

A total of 298 studies addressing SLR, SLR update (41 studies) and LSR (257 studies) were retrieved during the search process. First, we excluded duplicated studies, and then we applied the IC and EC on the title, abstract and keywords of these studies. As a result, we selected 17 candidate studies in this stage. Next, we read and applied the IC and EC on the full text of the candidate studies, and selected 12 studies in this stage. We also performed an iteration of the backward and forward snowballing \cite{wohlin16} to identify additional studies through the list of references and citations of the selected studies. As a result, we selected three more studies. Thus, a total of 15 studies compose our final set of studies. Table \ref{table:studies} compiles in a list all 18 studies (S1 -- S18), selected for the following stages of the meta-ethnography process execution. It includes our final set of 15 selected studies categorized by their main topic and the books and books' chapter previously selected (+3).

\definecolor{light-gray}{gray}{0.95} 

\begin{table}[ht!]
\footnotesize
\centering
\caption{Selected studies for the next stages of the meta-ethnography}
\begin{tabular}{llcc}
\toprule
\textbf{ID}  & \textbf{Main topic}   & \textbf{Publication Year} & \textbf{Reference}  \\
\midrule
S1  & SLR update   & 2020             & \cite{Watanabe20}     \\ 
\rowcolor{light-gray}
S2  & SLR update   & 2020             & \cite{Wohlin2020}     \\ 
S3  & SLR update   & 2020             & \cite{Mendes2020}     \\ 
\rowcolor{light-gray}
S4  & SLR update   & 2020             & \cite{Felizardo20}    \\ 
S5  & SLR update   & 2019             & \cite{Nepomuceno2019} \\ 
\rowcolor{light-gray}
S6  & SLR update   & 2018             & \cite{Felizardo18} \\ 
S7  & SLR update   & 2017             & \cite{Garces17}       \\ 
\rowcolor{light-gray}
S8  & SLR update   & 2016             & \cite{felizardo16} \\ 
S9  & SLR update   & 2008             & \cite{Dieste08a}   \\ 
\rowcolor{light-gray}
S10 & SLR process  & 2017             & \cite{Kuhrmann2017}   \\ 
S11 & SLR process  & 2015             & \cite{Kitchenham15}   \\ 
\rowcolor{light-gray}
S12 & SLR process  & 2007             & \cite{Kitchenham07}   \\ 
S13 & LSR          & 2022             & \cite{Simmonds2022}   \\ 
\rowcolor{light-gray}
S14 & LSR          & 2019             & \cite{Cochrane2019}\\ 
S15 & LSR          & 2017             & \cite{Elliott17} \\ 
\rowcolor{light-gray}
S16 & DevOps       & 2015             & \cite{Bass2015}       \\ 
S17 & DevOps       & 2010             & \cite{Humble2010}     \\ 
\rowcolor{light-gray}
S18 & Open Science & 2020             & \cite{Mendez2020}     \\ 
\bottomrule
\end{tabular}
\label{table:studies}
\end{table}

\subsubsection{Reading the studies}

During the third stage, the authors read the set of selected studies and perform the data extraction \cite{Noblit/1988}. In the context of this study, all the authors are familiar with the SLR process and the software development process. Therefore, the authors' experience and their interactions facilitated the understating of the content of the selected studies. The first author extracted the data using a data extraction form, which was revised by the second author.

The data extraction form was built to obtain (i) the general objective of the study; (ii) its main results and contributions; (iii) if the study proposes an approach or the use of a technique, a description of the approach or technique presented and whether it has been validated; and (iv) if the study presented a process, the description of each process activity including roles, inputs, processing, and outputs. 

Analyzing the 18 selected studies, as shown in Table \ref{table:studies}, the majority of the selected studies address the SLR Update process in SE since our goal was to mitigate intermittent SLR update issues in SE, which requires a deep understanding on the advancements on SLR updates in SE over the past years. Notably, SLR updates have gained the attention of the SE community. In 2020 four studies proposing improvements in conducting SLR updates in SE were published (S1, S2, S3 and S4). Watanabe \textit{et al.}  \cite{Watanabe20} (S1) propose and evaluate the use of text classification to provide automated support in the SLR update studies selection activity. Wohlin \textit{et al.}  \cite{Wohlin2020} (S2) propose guidelines on the search strategy to update SLRs in SE. Mendes \textit{et al.}  \cite{Mendes2020} (S3) recommend using a decision framework on when to update SLRs in SE. Felizardo \textit{et al.}  \cite{Felizardo20} (S4) present an experience report on how to transfer the know-how of SLRs to facilitate their updates through the instantiation of a knowledge management model.

Other five studies addressing SLR updates were selected (S5, S6, S7, S8 and S9). Besides S4, two other experience reports on SLR updates were considered in our study (S5 and S6). Nepomuceno \& Soares \cite{Nepomuceno2019} (S5) present a systematic mapping and survey on how researchers are evolving their SLRs and what they think about SLR updates. Garcés \textit{et al.} \cite{Garces17} (S7) relate the authors' experience in updating two SLRs using automated techniques based on VTM (Visual Text Mining). Felizardo \textit{et al.} \cite{felizardo16} (S8) introduced the adoption of forward snowballing \cite{wohlin16} to search for studies to update SLRs in SE. Two years later, Felizardo \textit{et al.}  \cite{Felizardo18} (S6) evaluated the use of different electronic databases for applying forward snowballing to update secondary studies. Study S2 presents a combined and more recent investigation addressing the approaches described in S5 and S6. Last but not least, Dieste \textit{et al.} \cite{Dieste08a} (S9) propose a process to perform SLR updates in SE, taking into account lessons learned from updating an SLR. 

Regarding the traditional SLR process, we considered the well-known Kitchenham \& Charters \cite{Kitchenham07} guidelines (S12) as well as Kitchenham's \textit{et al.} Book \cite{Kitchenham15} (S11) which contains an extended description of the SLRs process in SE as well as the update of the guidelines proposed in S12. Kuhrmann \textit{\textit{et al.}} \cite{Kuhrmann2017} (S10) present an experience-based guideline to aid researchers in designing SLRs in SE with emphasis on the studies' search and selection procedures.

LSRs were introduced in the medical field in 2017 by Elliott \textit{et al.} \cite{Elliott17} (S15), aiming to incorporate relevant new evidence to SLRs as it becomes available. Two years later, the Cochrane DL embraced the LSR concept and published its guidelines for the production and publication of Cochrane LSRs \cite{Cochrane2019} (S14). More recently, Simonmds \textit{et al.}  \cite{Simmonds2022} (S13) described the general principles of LSR in a book chapter, when they might be of particular value, and how its procedure differs from conventional SLRs.

The three last studies included in our analysis are the book of Bass \textit{et al.} \cite{Bass2015} (S16), which describes the DevOps concept through a software architecture perspective detailing each step of the DevOps process. The book of Humble \textit{et al.}  \cite{Humble2010} (S17) which presents the whole DevOps, explaining in detail the CI/CD practices. Finally, the recently published book chapter by Mendez \textit{et al.}  \cite{Mendez2020} (S18) on OS for SE includes the OS definition, why SE researchers should engage in it, and how they should do it. 



\subsubsection{Determining how the studies are related}

In this fourth stage, using the extracted data from the stage (3) and revisiting the select studies when needed, we extracted the metaphors of each selected study. These metaphors were phrases, ideas, and concepts that could be relevant to detecting relationships or connections among the studies. Next, a list describing all the identified relationships among the studies is created. We used spreadsheets to support this step. When we started to extract the relationships between studies, we noticed that all studies have some process factor that directly impacts the updating of an SLR. We summarize these relationships in Table \ref{table:metaphors}. The process factors are described in the first column of Table \ref{table:metaphors}. 


\begin{table}[ht!]
\footnotesize
\caption{Relationships among the selected studies}
\centering
\begin{tabular}{p{2cm}p{9cm}}
\toprule
\textbf{Process factor}  & \textbf{Related studies}     \\
\midrule
\textbf{Planning }  & S3, S4, S5, S7, S9, S10, S11, S12, S13, S14, S15, S16, S17     \\ 
Study selection     & S1     \\ 
Search strategy    & S2, S6, S8     \\ 
\rowcolor{light-gray}
\textbf{Executing}    & S1, S2, S3, S4, S5, S6, S7, S9, S10, S11, S12, S13, S14, S15, S16, S17      \\ 
\textbf{Reporting}  & S4, S5, S9, S10, S11, S12, S3, S14, S15, S16, S17, S18    \\ 
\rowcolor{light-gray}
\textbf{Monitoring}    & S13, S14, S15, S16, S17    \\ 
\bottomrule
\end{tabular}
\label{table:metaphors}
\end{table}


As shown in Table \ref{table:metaphors}, we detected studies related to specific SLR update planning factors: study selection and search strategy. 
Despite the base of the SLR update process being the original SLR process (planning, executing and reporting), i.e. re-execute (and adapt, if necessary) the original review protocol, we identified that the DevOps concepts and process could be seen as a metaphor for building a process of continuous integration and delivery of evidence to support keeping SLRs up-to-date. Monitoring, a DevOps process factor (S16-S17), is connected to the LSR studies (S13-S14-S15) which are connected to the other three SLR and SLR update factors (planning, executing and reporting). In addition, the OS study (S18) is connected to the reporting factor since it addresses the availability of information. 

\subsubsection{Translating the studies into one another}

The main goal of this stage is to compare the metaphors (text fragments) extracted from the studies. Unfortunately, Noblit and Hare \cite{Noblit/1988} do not describe how to perform this step, but one suggestion is to compare the synthesis of each study progressively \cite{Silva13}. 

\newcounter{mpFootnoteValueSaver}
\setcounter{mpFootnoteValueSaver}{\value{footnote}}
\stepcounter{mpFootnoteValueSaver}
\footnotetext[\value{mpFootnoteValueSaver}]{\url{https://doi.org/10.5281/zenodo.6503143}}

We analyzed the metaphors observed in stage (4), comparing them and analysing their relationships. To construct the relations among the studies, we opted to guide this construction based on the original SLR process activities as demonstrated in Table \ref{table:metaphors} because it is also the basis process for updating an SLR. In addition, the LSR process follows the same base structure as the SLR process. In this sense, first, we build a table based on the elements of the relationships detected in stage (4). Second, we separated each metaphor into (i) general metaphors: fragments that address the relationship among the studies and procedural elements generally; (ii) specific metaphors: fragments that address specific relations among each activity necessary to perform an SLR update. Finally, we checked our list of metaphors to see if any of them directly mitigated the SLR intermittent update issues. Therefore, we checked if the relationships were found to converge to the objective of our study. We also added the references that support our findings\footnotemark[\value{mpFootnoteValueSaver}]. 



Analyzing the identified relationships, it is clear that the selected studies' practices, processes, activities, and recommendations are strongly interconnected. Hence, the systematization of the evidence found in this step in the form of a process can contribute to the establishment of a dedicated process for mitigating the intermittent SLR update problem. 

\subsubsection{Synthesizing the translations}

\begin{figure} [ht!]
    \centering
    \includegraphics [width=1\linewidth]{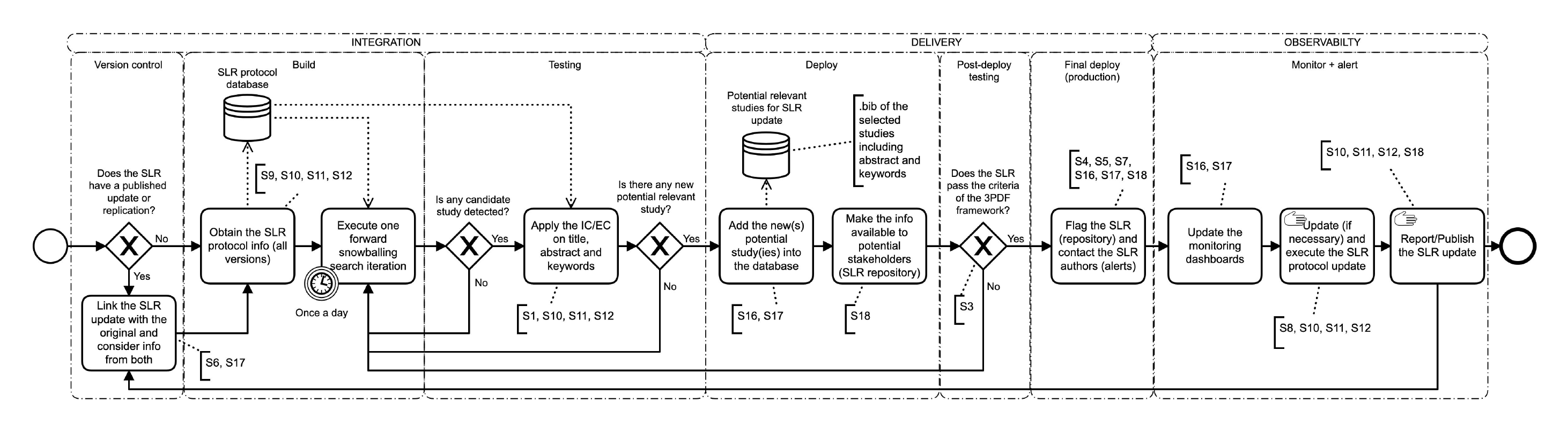}
    \caption{CSLR process}
    \label{CSLR1}
\end{figure}

During the sixth stage, we constructed a translation synthesis. The translations of studies result in many metaphors. They are compared to verify similarities and/or if some metaphors can encompass others. The result of this stage is usually represented as diagrams or figures \cite{Noblit/1988}. 
In the context of our study, we compared and systematized the translations mapped in step (5) as a continuous process for assessing new evidence and evaluating the need of update SLRs (Continuous Systematic Review -- CSLR). Figure \ref{CSLR1} illustrates the three stages of the CSLR process (Integration, Delivery and Observability) with their phases and activities. The CSLR BPMN diagram is also available online\footnotemark.

The CSLR process starts with the \textit{Integration} stage and \textit{Version Control} phase (see Figure \ref{CSLR1}). In this phase, the first activity is to verify if the SLR has an update or replication published. If yes, these studies must be linked and considered in the next process activity. 

In the second phase of the \textit{Integration} stage, the \textit{Build} phase, the first activity is to obtain the protocol information of all considered studies (original SLR, updates, replications -- if it exists) and store them in a database. These protocols information include: Addressed research questions, the period covered by the SLR execution, list of included studies, inclusion and exclusion criteria, and quality criteria (if adopted). 

In the next activity of the \textit{Build} phase, using the original SLR and its list of included studies, one iteration of the forward snowballing using the Google Scholar as DL is performed \cite{Wohlin2020} once a day. If there is an SLR update and/or replications linked to this SLR, these studies and their list of included studies must be considered in the snowballing execution. 

The last phase of the \textit{Integration} stage, the \textit{Testing} phase, starts checking if any study was detected by the snowballing process (candidate studies). If no study was detected, the process returns to forward snowballing execution activity. If one or more studies were detected, the IC and EC criteria must be applied to the candidate studies' title, abstract, and keywords. The study of Watanabe \textit{et al.} \cite{Watanabe20} proposed an automated alternative for this step using text classification techniques. The next activity is to verify if there is any potential relevant study according to the IC and EC. If not, the process returns to the activity of snowballing execution. If yes, the process moves to the \textit{Delivery} stage. 

The \textit{Delivery} stage starts with the \textit{Deploy} phase. The first activity in this phase is to add the new potential relevant studies to the database. In order to deal with automation \cite{Watanabe20} and to allow using reference management tools (such as Jabref and Zotero), the studies should be stored in the database in .bib format. The next activity is to make these potential relevant studies available to potential stakeholders. For that, we suggested uploading this information into a public repository such as Zenodo or ArXiv \cite{Mendez2020}. 

The next phase, the \textit{Post-deploy testing}, has one activity: apply the 3PDF framework proposed in \cite{Mendes2020} to analyse if the SLR needs to be updated. The 3PDF framework consists in answering 7 questions (steps) as described in \cite{Mendes2020}.
If the results of the 3PDF show that the SLR is suitable for an update, in the \textit{Final deploy} phase the SLR must be flagged as ``outdated'' and the authors of the original SLR, update and/or replication (if it exists) will be contacted. Otherwise, the process returns to the activity of defining the snowballing forward frequency. 

The last stage of the CSLR process is the \textit{Observability} stage. This stage has an unique phase: \textit{Monitor + Alert}. The two first activities of this stage are conditioned to process automation, i.e., the development of a dedicated SLR repository. It consists of updating the monitoring dashboards that report the SLR update data gathered during the CSLR process execution. 
The last two activities consist in performing the SLR update. It includes updating protocol items (if necessary) \cite{Mendes2020} and executing it; and reporting the results of the SLR update (e.g., publishing) \cite{Kitchenham15}. Finally, with the first process cycle finished, the new and completed SLR updated must be linked to its other version(s), re-starting the CSLR process from its beginning. 

\subsubsection{Expressing the synthesis}

In this last stage, the synthesis findings are disseminated to interested parties. In our case this concerns making the CSLR process available as a direction for future research on continuously keeping SLRs in SE up-to-date.


\section {Applying CSLR to a Published SLR}
\label{sec:validation}

To evaluate the CSLR process, we selected a suitable SLR as an instrument to execute the process through the conduction of a case study. Our goal is to evaluate the feasibility of applying the proposed CSLR process and observe its contributions to mitigating the SLR intermittent update issues. We translated our case study goal into two Research Questions (RQs): 
\textbf{RQ1:} \textit{How do the steps of the proposed CSLR process perform in practice?}
\textbf{RQ2:} \textit{Can the CSLR process help to mitigate the intermittent SLR update issue in SE?}

We follow the five main steps for conducting case studies proposed by \cite{Runeson12}: Design, preparation, collecting data, analysis and reporting. These steps are described hereafter. 

\subsection{Design}
\label{subsec:Design}

Our design consists in selecting an SLR to be the instrument (input) of the CSLR process and then executing the CSLR process steps manually.

We chose the SLR \cite{Kitchenham07a} which addresses the topic of cross-company vs within-company effort estimation, for the following reasons: (i) The SLR is published as a conference paper \cite{Kitchenham06}, and after as a journal paper \cite{Kitchenham07a} in renowned SE venues;  (ii) The SLR has being used as evaluation instrument by several other studies \cite{Wohlin2020, Mendes2020, felizardo16, wohlin16}; (iii) The SLR was last updated in 2014 (over seven years ago)\cite{Mendes14} -- with this, we can also evaluate the CSLR process when the SLR already has a published update; and (iv) The SLR update has as co-author one of the authors of this study.

\subsection{Preparation}
\label{subsec:Preparation}

We distributed the conduction of the case study between the first and last authors of this study. Both authors have experience in conducting SLRs and updates. The last author is co-author of the SLR Update \cite{Mendes14}.
According to SLR experience reports \cite{Felizardo20, Nepomuceno2019, Garces17}, the participation of a member who already participated in the previous review can facilitate the update process besides contribute to avoiding bias.


\subsection{Collecting data}
\label{subsec:Collecting data}

Data collection is based on the SLR chosen for the CSLR process evaluation. The CSLR process has several data collection activities that must be carried out according to the process execution (e.g. check if there is a published SLR update, obtain the list of included studies). 

\subsection{Analysis}
\label{subsec:Analysis}

We report our analysis on how the process steps perform in practice by describing the case of applying each CSLR process activity to the selected SLR. 

The first activity of CSLR involves verifying if our SLR candidate has a published update. For that, we checked the citations of the SLR on Google Scholar (428 citations). As a result, we identified an update \cite{Mendes14} (SLR-Update) published in 2014, as well as two other studies published in 2016 that replicated the SLR update investigating different search strategies \cite{wohlin16} (SLR-UR1) and \cite{felizardo16} (SLR-UR2). Therefore, SLR-Update, SLR-UR1 and SLR-UR2 were selected in the version control phase (\textit{Integration} stage) of the CSLR process execution. It is important to mention that these same three studies were also identified in \cite{Wohlin2020} that used this same SLR as an investigation instrument. 

The next phase of the CSLR process (\textit{Build}) begins with obtaining protocol information from the studies selected in the version control phase. With the support of a spreadsheet, we extracted the following information from the original SLR, SLR-Update, SLR-UR1 and SLR-UR2: \textit{(i) Research questions }-- all studies investigated the same research questions;  \textit{(ii) IC, EC and quality criteria} -- the IC, EC and quality criteria were the same for all studies;  
\textit {(iv) Search strategy} -- the original SLR performed automated search on six SE Digital Libraries (DLs), 
manual search on individual journals and conference proceedings and reference checking (a.k.a backward snowballing). The SLR-Update used the same method as the original SLR except for adding Scopus as an extra DL and not performing a manual search. SLR-UR1 adopted backward and forward Snowballing \cite{wohlin16} and SLR-UR2 only forward snowballing; \textit{(v) Search strategy coverage period} -- the original SLR covered the period from 1990 to November 2006, the SLR-Update from December 2006 to end 2013 and SLR-UR1 and SLR-UR2 both from 2006 to 2013; and \textit{(vi) List of included studies} -- The original SLR included ten primary studies, the SLR-Update 11 additional primary studies. The SLR-UR1 did not include two studies included in the SLR-Update, but it included three other studies that did not include in the SLR-Update. The SLR-UR2 included all the studies included in the SLR-Update except for one study, and it also identified two other studies. Considering this scenario, for the CSLR, we considered all unique included studies from the original SLR, SLR-Update, SLR-UR1 and SLR-UR2, totalling a final list with 25 included studies published by the end of 2013. 


The second activity in the \textit{Build} phase is to execute one interaction of forward snowballing search technique on Google Scholar to identify new potential relevant studies. Using the 25 included studies and the original SLR, SLR-Update and SLR-UR1 and SLRU-R2 as seed, we performed an interaction of the forward snowballing technique we obtained a total of 2392 returned studies. Since the SLR-Update and both replications covered the search until 2014, we limited our search results from 2014 to February 2022, resulting in 858 returned studies. We exported the bibliographical data, including keywords and abstracts of all studies in .bib and .csv format.

Moving to the \textit{Testing} phase, the first step is to perform the initial cleaning of our set of returned studies. Since this process has been executed manually, we used the .csv file to remove duplicated studies and conference announcements. Thus, we arrived at a list of 444 unique candidate studies. Next, we applied the inclusion and exclusion criteria on the title, abstract and keywords of each study arriving in a set of 24 potential candidate studies. It is worth mentioning that all 24 studies contain at least one or more keywords that would allow an automated selection method such as \cite{Watanabe20} to identify these studies. The first and last author carefully analyzed the title, abstract and keywords of each study, and in a synchronous meeting, they decided through consensus what studies have a strong potential for inclusion. As a result, five new studies showed significant potential to be included in the SLR, and other five studies presented a new trend of investigations using cross-company and within-company mixed together to estimate software project effort (which could lead to updating the SLR protocol, including its research questions, to properly consider this new identified trend). The list with the 24 potential candidate studies and the ten selected potential studies to be included in a new update is available online\footnotemark[\value{mpFootnoteValueSaver}].

The \textit{Deploy} phase starts by adding the potential selected studies into a database and next making them available online in a repository. Since we are performing both activities manually, we created a .bib file with the ten potential included studies. We made them available at Zenodo\footnotemark[\value{mpFootnoteValueSaver}], an open dissemination research data repository.

The \textit{Post-deploy testing} phase consists of using the 3PDF framework \cite{Mendes2020} to verify if the SLR needs to be updated. Thus, we performed the seven steps (questions) of the 3PDF method. The answer to each question is available online\footnotemark[\value{mpFootnoteValueSaver}]. As a result, the SLR needs to be updated.

In the \textit{Delivery} stage, during the final deployment, we must flag the SLR with the status ``SLR suitable for update'' and communicate the SLR and SLR-Update (if existis) authors about the findings. As this is a feasibility study, we did not contact anyone with regard to these results yet.

The last stage of the CSLR process is the \textit{Observability} stage. The first activity described in the \textit{Monitor + alert} phase, flag the SLR and contact the authors, is not suitable for performing manually. The activities of performing and publishing the SLR update can be performed without these activities. However, updating the SLR is out of the scope of this study. Hence, since the SLR is not updated yet, the process returns to the \textit{Build} phase to search for new potential relevant studies.
 
\subsection{Reporting}
\label{subsec:Reporting}

\textbf{RQ1:} \textit{How do the steps of the proposed CSLR process perform in practice?} 

In general, the CSLR process flow seemed to be coherent. The stages, phases and activities were manually applicable in practice, addressing relevant aspects to help mitigating intermittent SLR update issues in SE. 

However, when it comes to automation aspects, while some activities are clear candidates for automation (e.g., the whole CSLR systematic surveillance and analysis of new potentially relevant studies), not all of them seem to be ready for complete automation yet. For instance, the application of some decision steps of the 3PDF framework, which still require human intervention and reasoning. Also, the update itself, including the rigorous full-text-based assessment, application of the quality assessment criteria, data extraction, and research synthesis, is seen as a manual activity to be conducted by researchers, as depicted in Figure \ref{CSLR1}.



\textbf{RQ2:} \textit{Can the CSLR process help to mitigate the intermittent SLR update issue in SE?}

During the execution of the case study, (manually) applying the CSLR process activities showed being feasible and enabled systematic surveillance and analysis of new potentially relevant studies. Furthermore, its application revealed the need to update an already published SLR. Hence, we conclude that the CSLR process can help to mitigate the  intermittent SLR update issue.

With proper automation (e.g., including SLR repositories, continuous integration facilities, and availability of information in an open and accessible way, monitoring dashboards), CSLR could help to systematically identify the need for updates and unnecessary SLR updates, as reported in \cite{Mendes2020}, could be avoided. Hence, we perceive making the CSLR process available as important to foster research in this direction.

\section{Discussion} 
\label{sec:discussion}

The motivation for the elaboration of the CSLR process is to provide a systematic process to help mitigating the intermittent SLR update problem contributing to avoiding missing new potential new relevant research in evidence-syntheses or decision-making. According to Nepomuceno \textit{et al.} \cite{Nepomuceno2019}, actions to keep SLR updated are of great importance to the SLR research field. 

The CSLR process unifies several pieces of knowledge on SE SLR updates that have been investigated separately, integrating them with DevOps and OS concepts. As a consequence, we observed benefits such as (i) facilitating the identification if the SLR has been updated or not (ii) assisting the identification (search and selection) of potentially relevant evidence; (iii) making potentially relevant evidence available in open repositories that are freely accessible by the SE community; (iv) supporting the decision on the need to update an SLR; and (v) supporting SLR authors throughout the update process. 

One may question the similarities between LSR and CSLR. Indeed, both have the same aim of keeping the SLR up to date \cite{Elliott17, Simmonds2022}. LSR is an medicine review approach to update an SLR continually (fully manually) requiring authors to make explicit commitments as to the frequency of search and screening execution as well as publication updates \cite{Cochraine2019a}. On the other hand, inspired in SE DevOps and OS practices, the CSLR process explores automation opportunities based on study findings investigated in the SE area on SE SLR updates (e.g. \cite{Wohlin2020, Mendes2020}) aiming to provide continuous and systematic surveillance and analysis of potential new relevant evidence. The need for an update is then decided by using the 3PDF framework. Furthermore, the idea of making the SLR protocol information and intermediate results openly available is not explored within the LSR context. LSR relates to continual SLR updates, while CSLR aims at continuous SLR updates.

In a more general view, the CSLR could also be an instrument to direct the SE community on research subjects that have been investigated -- if an SLR is often cited, it gives evidence that the subject of study addressed by it is constantly evolving. This fact leads to questions such as: Does the SLR remain relevant? Is the SLR up to date? Does it needs to be updated? -- As observed in our case study, new research trends can be identified, leading to the proposition of other research directions (questions) on a research subject. 

Besides, the CSLR process showed to be effective during our evaluation and opens avenues for automating its activities and pipeline. Researchers have investigated automation of the SLR process over the years \cite{Felizardo2020Book}. However, to the best of our knowledge, only two studies are addressing automation alternatives for SLR updates \cite{Felizardo14, Watanabe20}. Moreover, both studies are focused only on the study selection activity. Therefore, there is a lack of approaches that automate or semi-automate other SLR update activities or approaches that integrate the update activities.

\section{Threats to Validity} 
\label{sec:threats}

\textbf{Construct validity.} We followed well-known guidance and advice on designing and conducting meta-ethnography studies \cite{Noblit/1988} and case studies in SE \cite{Runeson12}. We also carefully selected the SLR to apply CSLR (cf. Section \ref{subsec:Design}).

\textbf{Internal validity and reliability.} Since meta-ethnography is an interpretive approach to synthesis, we addressed the validity and reliability of our synthesis by performing discussions among the authors during the conduction of all seven steps of the method, and we evaluated the outcome (CSLR process) through a case study. Regarding the case study, it was conducted by the first and last author of this study. The last author is co-author of the SLR Update \cite{Mendes14} to which CSLR was applied, which avoids misunderstandings with regard to the SLR protocol \cite{Nepomuceno2019}. 

\textbf{External validity.} Similar case studies could have been applied to more SLRs to improve the generalizability of our results. However, manually applying the process involves significant effort. Nevertheless, we intend to extend this study by addressing more SE SLRs, also counting on the participation of external researchers to investigate the CSLR process. 

\section{Conclusion}
\label{sec:conclusion}

In this study, we proposed and evaluated the Continuous Systematic Literature Review (CSLR) concept and process to support SLRs updates in SE. We structured the CSLR process by synthesizing evidence through a meta-ethnography integrating knowledge from varied research areas. 


The CSLR concept and process comprises a continuous and systematic surveillance and analysis of potential new relevant evidence for published SLRs in a timely manner, contributing to keeping SLRs up to date. 
The results of our evaluation suggest that CSLR can help to mitigate the relevant SLR intermittent update issue for SLRs in SE. Besides, the CLSR process provides a systematic pipeline towards automating and managing SLR updates activities.

Hence, we believe that making the CSLR process available can foster future work investigating more effective ways for continuously summarizing existing SE evidence. For instance, further evaluating the CSLR process and investigating automating its process pipeline.

\bibliographystyle{unsrt}  
\bibliography{references}  

\end{document}